\documentclass[12pt]{article}
\usepackage{subeqn}
\usepackage{epsfig}
\newcommand{\be}{\begin{equation}}
\newcommand{\ee}{\end{equation}}

\newcommand{\no}{\noindent}
\newcommand{\ce}{\begin{center}}
\newcommand{\nc}{\end{center}}

\makeatletter
\@addtoreset{equation}{section}
\makeatother

\def\sqr#1#2{{\vcenter{\vbox{\hrule height.#2pt
\hbox{\vrule width.#2pt height#1pt \kern#1pt
\vrule width.#2pt} \hrule height.#2pt}}}}

\def\operp{\hbox{${\kern+.25em{\bigcirc}
\kern-.85em\bot\kern+.85em\kern-.25em}$}}

\def\lsim{\;\raise0.3ex\hbox{$<$\kern-0.75em\raise-1.1ex\hbox{$\sim$}}\;}
\def\gsim{\;\raise0.3ex\hbox{$>$\kern-0.75em\raise-1.1ex\hbox{$\sim$}}\;}
\def\no{\noindent}

\def\ce{\centerline}
\def\ve{\vfill\eject}
\def\rdots{\mathinner{\mkern1mu\raise1pt\vbox{\kern7pt\hbox{.}}\mkern2mu
\raise4pt\hbox{.}\mkern2mu\raise7pt\hbox{.}\mkern1mu}}

\def\e e{$e^+ e^-$ }

\begin{document}

\ce{\bf TREFOIL SOLITONS, ELEMENTARY FERMIONS, AND $SU_q(2)$}

\vskip.3cm

\ce{\it Robert J. Finkelstein}
\vskip.3cm

\ce{Department of Physics and Astronomy}
\ce{University of California, Los Angeles, CA 90095-1547}

\vskip1.0cm

\no{\bf Abstract.} By utilizing the gauge invariance of the $SU_q(2)$
algebra we sharpen the basis of the $q$-knot phenomenology.

\ve

\section{Introduction.}

In constructing a knot model of the elementary particles the primary
challenge is to establish a procedure for correlating the knots with
the particles.  In earlier work$^{1,2}$ we have attempted to establish
this map semi-empirically.  With the aid of the quantum group $SU_q(2)$
we would now like to describe a clear set of {\it a priori} rules for
mapping the knots onto the particles.

The quantum group $SU_q(2)$ offers a possible realistic implementation
of a model of elementary particles as knotted flux tubes, since on the
one hand the linearization of $SU_q(2)$ approximates in low order the
symmetry group of standard electroweak,$^{1,2}$ while on the other
hand $SU_q(2)$ underlies the description of knots.$^3$  In this model
the simplest particles (elementary fermions) are the simplest knots
(trefoils).  There are four families of elementary fermions and there 
are indeed four trefoils.  If it is possible to match the four
families with the four trefoils one may then ask whether the three
members of each family are the three lowest states of excitation of a
vibrating trefoil.  Since all members of a family have the same
quantum numbers, $(t,t_3,t_0)$ representing isotopic spin, its 
third component, and the hypercharge respectively, while each trefoil
is also characterized by three integers $(N,w,r)$ representing the
number of crossings, the writhe and the rotation, respectively, it is
necessary to establish a correspondence between $(N,w,r)$ and $(t,t_3,t_0)$.  This correspondence may be established by introducing the
$D^j_{mm^\prime}$, the irreducible representations of $SU_q(2)$ and
labelling both the trefoils and the elementary fermions by the same
$D^j_{mm^\prime}$.  In order to utilize half-integer representations
of $SU_q(2)$ and also to respect the knot constraint that requires a
difference in parity between $w$ and $r$ we choose
\be
\begin{array}{rcl}
j &=& \frac{N}{2} \\
m &=& \frac{w}{2} \\
m^\prime &=& \frac{r+1}{2} \\
\end{array}
\ee
\no Then the knots will be labelled by $D^{N/2}_{\frac{w}{2}\frac{r+1}{2}}$.

It remains to relate the $(t,t_3,t_0)$ to the $(j,m,m^\prime)$.

\vskip.5cm

\section{Correspondence of the Four Families with the Four Trefoils.}

The $(2j+1)$-dimensional irreducible representations of $SU_q(2)$ are
\be
D^j_{mm^\prime}(a,\bar a,b,\bar b) = \Delta^j_{mm^\prime} \sum_{s,t}
\left\langle\matrix{n_+ \cr s \cr}\right\rangle_1
\left\langle\matrix{n_- \cr t}\right\rangle_1
q_1^{t(n_++1-s)}(-1)^t\delta(s+t,n^\prime_+) 
a^sb^{n_+-s}\bar b^t\bar a^{n_--t} 
\ee
\no where
\be
\begin{array}{rcl}
n_\pm &=& j\pm m \\ n^\prime_\pm &=& j\pm m^\prime \\
\end{array} \quad
\left\langle\matrix{n \cr s \cr}\right\rangle_1 =
{\langle n\rangle_1!\over \langle s\rangle_1!\langle n-s\rangle_1!}
\quad \langle n\rangle_1 = {q_1^{2n}-1\over q_1^2-1} \nonumber
\ee
\no and  
\be
\Delta^j_{mm^\prime} = \left[{\langle n^\prime_+\rangle_1!~
\langle n^\prime_-\rangle_1!\over
\langle n_+\rangle_1!~\langle n_-\rangle_1!}\right]^{1/2} \qquad
q_1 = q^{-1} 
\ee
\no Here the arguments $(a,b,\bar a,\bar b)$ obey the following
algebra:
\begin{center}
\begin{tabular}{lllr}
$ab= qba$ & $a\bar a+b\bar b = 1$ & $b\bar b = \bar bb$ & \\
$a\bar b =q\bar ba$ & $\bar aa + q_1^2\bar bb = 1$ & & \hspace{1.5cm}(A)
\end{tabular}
\end{center}
\no In the limit $q=1$, where $\langle n\rangle_1\to n$ and the
arguments commute, the $D^j_{mm^\prime}$ become the irreducible
representations of $SU(2)$.  According to (2.1) the labelling of each
trefoil by $D^{3/2}_{\frac{w}{2}\frac{r+1}{2}}$ leads to the 
monomials shown in Table 1.
\vskip.3cm
\begin{center}
{\bf Table 1.}
\end{center}
\be
\begin{array}{ccc}
\underline{(w,r)} & ~~~\underline{D^{3/2}_{\frac{w}{2}\frac{r+1}{2}}}
& ~~~\underline{\bar D^{3/2}_{\frac{w}{2}\frac{r+1}{2}}} \\
(-3,2) & ~~~D^{3/2}_{-\frac{3}{2}\frac{3}{2}}\sim\bar b^3 & ~~~
\sim b^3 \\
(3,2) & ~~~D^{3/2}_{\frac{3}{2}\frac{3}{2}}\sim a^3 & ~~~ 
\sim \bar a^3 \\
(3,-2) & ~~~ D^{3/2}_{\frac{3}{2}-\frac{1}{2}}\sim ab^2 & ~~~
\sim\bar b^2\bar a \\
(-3,-2) & ~~~ D^{3/2}_{-\frac{3}{2}-\frac{1}{2}}\sim\bar a^2\bar b &
~~~\sim ba^2 \\
\end{array}
\ee
\no where the numerical coefficients appearing in $D^{3/2}_{mm^\prime}$
have been dropped.  The corresponding labelling of the particles is
accomplished first by attaching the $D^{N/2}_{mm^\prime}(a,\bar a,b,
\bar b)$ to the normal modes representing momentum and spin.  Then the
quantum fields will lie in the (A) algebra.  We shall assume that the
antiparticle fields are the adjoint fields in this algebra (as well
as the Dirac conjugate fields in the usual way).  Therefore one must
attach $\bar D^{N/2}_{mm^\prime}$ to the normal modes of the usual
antiparticle field.

Since the $D^j_{mm^\prime}$ are assumed to mediate the correspondence
between the knots and the particles, we next assign $(t,t_3,t_0)$
as well as $(N,w,r)$ to $D^j_{mm^\prime}$, and since the fermions
all have $t=1/2$ and $t_3=\pm 1/2$ while the trefoils have $N=3$ and
$w=\pm 3$, we set
\begin{eqnarray}
t &=& \frac{N}{6} \\
t_3 &=& -\frac{w}{6}
\end{eqnarray}

To find the third relation for $t_0$, or for $Q=t_3+t_0$, we must 
first require that the particle and antiparticle have opposite charge
$Q$.  Next note that every term of (2.1) contains a product of 
non-commuting factors that may be reduced (after dropping numerical
factors) to the form
\be
a^{n_a}\bar a^{n_{\bar a}}b^{n_b}\bar b^{n_{\bar b}}
\ee
\no where $n_a = s$ and $n_{\bar a}=n_--t$.

But the factor $\delta(s+t,n^\prime_+)$ appearing in (2.1) implies
\[
n_+^\prime = n_a + (n_--n_{\bar a})
\]
\no or
\be
n_a-n_{\bar a} = m+m^\prime
\ee
\no Note that (2.8) holds for \underline{all} terms and \underline{all}
representations.

Since particles and antiparticles have adjoint symbols as well as
opposite charge $(Q)$ we may set
\be
Q = k(n_a-n_{\bar a})
\ee
\no or by (2.8)
\be
Q = k(m+m^\prime)
\ee

According to (2.10) one sees that $k=-1/3$ is the only choice of $k$
that agrees with the pairs $(w,r)$ of the trefoils and the charges
of the four families as shown in Table 2.
\ve

\begin{center}
{\bf Table 2.}
\end{center}
\be
\begin{array}{ccrc}
\underline{(w,r)} \quad & \underline{D^{3/2}_{\frac{w}{2}\frac{r+1}{2}}} \quad & \underline{Q} \quad & \underline{\rm Family} \\
(-3,2) \quad & D^{3/2}_{-\frac{3}{2}\frac{3}{2}} \quad & 0
\quad & (\nu_e,\nu_\mu,\nu_\tau) \\
(3,2) \quad & D^{3/2}_{\frac{3}{2}\frac{3}{2}} \quad & -1 \quad &
(e,\mu,\tau) \\
(3,-2) \quad & D^{3/2}_{\frac{3}{2}-\frac{1}{2}} \quad & -\frac{1}{3}
\quad & (dsb) \\
(-3,-2) \quad & D^{3/2}_{-\frac{3}{2}-\frac{1}{2}} \quad & \frac{2}{3}
\quad & (uct) \\
\end{array}
\ee
\no Hence $k=-1/3$ and
\begin{eqnarray}
Q &=& -\frac{1}{3} (n_a-n_{\bar a}) \\
Q &=& -\frac{1}{3} (m+m^\prime)
\end{eqnarray}

It follows also from (2.9) that $Q=0$ if
\be
n_a=n_{\bar a} 
\ee
\no In this case $a$ and $\bar a$ may be eliminated from (2.7)
in favor of $b$ and $\bar b$ since
\be
a^n\bar a^n = \prod^{n-1}_{s=0} (1-q^{2s}b\bar b)
\ee
\no by (A).  Therefore neutral states (neutrinos and neutral bosons)
lie entirely in the $(b,\bar b)$ subalgebra.

In knot coordinates $(N,w,r)$ the expression (2.13) for the charge
of the fermions becomes
\be
Q = -\frac{1}{6} (w+r+1)
\ee
by (1.1).  If one compares with the standard expression for $Q$
\be
Q = t_3+t_0
\ee
\no then by (2.6) and (2.16)
\begin{subequations}
\be
t_3 = -\frac{w}{6}
\ee
\be
t_0 = -\frac{1}{6} (r+1)
\ee
\end{subequations}
\no so that we have Table 3 agreeing with (2.18) and the
usual assignments of $Q$, $t_3$, and $t_0$
\vskip.3cm
\begin{center}
{\bf Table 3.}
\end{center}
\be
\begin{array}{crrr}
\underline{(w,r)} \qquad & \underline{Q} \qquad & \underline{t_3}
\qquad & \underline{t_0} \\
(-3,2) \qquad & 0 \qquad & \frac{1}{2} \qquad & -\frac{1}{2} \\
(3,2) \qquad & -1 \qquad & -\frac{1}{2} \qquad & -\frac{1}{2} \\
(3,-2) \qquad & -\frac{1}{3} \qquad & -\frac{1}{2} \qquad & 
\frac{1}{6} \\
(-3,-2) \qquad & \frac{2}{3} \qquad & \frac{1}{2} \qquad & \frac{1}{6} \\
\end{array}
\ee

The fact that there are two charges in the standard electroweak theory
corresponding to the separate groups $SU(2)$ and $U(1)$, should be
evident as well in the knot model as implemented by $SU_q(2)$.
Therefore let us consider
\be
Q_b = n_b-n_{\bar b}
\ee
\no where
\be
n_b = n_+-s \quad \mbox{and} \quad n_{\bar b}=t
\ee
\no Then the factor $\delta(s+t,n_+^\prime)$ in (2.1) implies
\be
(n_+-n_b) + n_{\bar b} = n_+^\prime
\ee
\no or
\be
m-m^\prime = n_b-n_{\bar b}
\ee
\no again holding for all terms and all representations.

We may now consider the two charges derived for $D^j_{mm^\prime}$,
namely
\begin{eqnarray}
Q_a &=& -\frac{1}{3} (n_a-n_{\bar a}) = -\frac{1}{3} (m+m^\prime) \\
Q_b &=& -\frac{1}{3} (n_b-n_{\bar b}) = -\frac{1}{3}(m-m^\prime)
\end{eqnarray}
\no By (1.1)
\begin{eqnarray}
Q_a &=& -\frac{1}{6}(w+r+1) \\
Q_b &=& -\frac{1}{6}(w-r-1)
\end{eqnarray}
\no By (2.26), (2.27), and (2.18)
\begin{eqnarray}
Q_a &=& -\frac{1}{3} (m+m^\prime) = t_3+t_0 \\
Q_b &=& -\frac{1}{3}(m-m^\prime) = t_3-t_0
\end{eqnarray}
\no By (2.28) and (2.29)
\begin{eqnarray}
m &=& -3t_3 \\
m^\prime &=& -3t_0 
\end{eqnarray}
\no in agreement with (2.19) and assignments of $D^{3/2}_{mm^\prime}$
in (2.11).

Ignoring their derivation, let us next test the same relations
(2.30) and (2.31) on the vector bosons.  Since the vectors are
responsible for pair production we shall represent them by
ditrefoils with $N=6$.  Then by (2.6) and (1.1)
\be
t = \frac{N}{6} = 1
\ee
\be
j = \frac{N}{2} = 3
\ee
\no By (2.30) and (2.31) one has Table 4.
\vskip.3cm
\begin{center}
{\bf Table 4.}
\end{center}
\be
\begin{array}{lcrcll}
& ~~~ \underline{t} ~~& \underline{t_3} ~~& \underline{t_0} &
\underline{D^3_{mm^\prime}} & \\
W^+ & ~~~1 ~~& 1 ~~& 0 ~~~& D^3_{-30} ~~~~~& \sim\bar b^3a^3 \\
W^- & ~~~1 ~~& -1 ~~& 0 ~~& D^3_{30} ~~~~~& \sim a^3b^3 \\
W^3 & ~~~1 ~~& 0 ~~& 0 ~~& D^3_{00} ~~~~~& \sim f_3(b\bar b) \\
W^0 & ~~~1 ~~& -1 ~~& 1 ~~& D^3_{3-3} ~~~~~& ~~b^6 \\
\end{array}
\ee
\no where $t_3$ and $t_0$ are taken from standard electroweak
theory.  Here
\be
\begin{array}{rcl}
f_3(b\bar b) &=& \prod^2_{s=0}(1-q^{2s}\bar bb)-q^2\langle 3\rangle_1^2
(b\bar b)\prod^1_{s=0}(1-q^{2s}b\bar b)+q_1^2\langle 3\rangle_1^2
(b\bar b)^2 \\
& &\times (1-b\bar b)^2(1-b\bar b)-q_1^{12}(b\bar b)^3 \\
\end{array}
\ee

The connection with knots is expressed in the trefoil case by
$D^{N/2}_{\frac{w}{2}\frac{r+1}{2}}$.  If the ditrefoil is comprised
of two connected trefoils, one has $D^{N/2}_{-3t_3,-3t_0}$ where
$t_3$ and $t_0$ are related to the knot parameters, $w$ and $r$, by 
Table 5.
\begin{table}
\begin{center}
{\bf Table 5.} \\
\vskip.3cm
\begin{tabular}{l|ccc}
&\underline{$W^+~\mbox{and}~ W^-$} & \underline{$W^3$} & \underline{$W^0$} \\
$t$ & $N/6$ &  $(N-1)/6$ &  $(N-1)/6$ \\
$t_3$ & $-w/6$ &  $w-1$ & $w$ \\
$Q$ & $-(1/3)r$ & $r$ & $r $\\
\end{tabular}
\end{center}
\end{table}

Table 5 agrees with Ref. 2 where the vector ditrefoil is
composed of two classically connected trefoils.  It would be
preferable to examine a ditrefoil as a quantum mechanical composite
of two trefoils.

\vskip.5cm

\section{Conservation Laws.}

The knot-particle interpretation may be related to standard
electroweak theory by ascribing additional structure to the normal
modes of the quantum fields.  Instead of point particles
$(t,t_3,t_0)$ we now associate solitonic knots $(N,w,r)$ with the
normal modes.  The solitonic knots display internal degrees of 
freedom and excited states that are defined by the $SU_q(2)$
algebra.  The gauge transformations on this algebra (A) that
correspond to the charges $Q_a$ and $Q_b$ are
\begin{subequations}
\be
\begin{array}{rcl}
a^\prime &=& e^{i\varphi_a} a \\
\bar a^\prime &=& e^{-i\varphi_a} \bar a \\
\end{array}
\ee
\be
\begin{array}{rcl}
b^\prime &=& e^{i\varphi_b} b \\
\bar b^\prime &=& e^{-i\varphi_b} \bar b \\
\end{array}
\ee
\end{subequations}
\no The algebra is invariant under these transformations.  The
induced gauge transformations on the solitonic modes are
\begin{subequations}
\be
\begin{array}{rcl}
U_a D^j_{mm^\prime} &=& e^{i\varphi_a(n_a-n_{\bar a})}
D^j_{mm^\prime} \\
&=& e^{-i3\varphi_aQ_a} D^j_{mm^\prime} \\
\end{array}
\ee
\no and
\be
\begin{array}{rcl}
U_b D^j_{mm^\prime} &=& e^{i\varphi_b(n_b-n_{\bar b})}
D^j_{mm^\prime} \\
&=& e^{-i3\varphi_bQ_b} D^j_{mm^\prime} \\
\end{array}
\ee
\end{subequations}

The standard action contains a trilinear term describing the
interaction of fermions and vector bosons.  Here that interaction
is multiplied by an additional factor of the following form:
\be
\bar F_1W_2F_3 \longrightarrow \bar D^{3/2}_{m_1m_1^\prime}
D^1_{m_2m_2^\prime} D^{3/2}_{m_3m_3^\prime}
\ee
\no Under the gauge transformations (3.1) the product (3.3) is by
(3.2) multiplied by
\begin{subequations}
\be
e^{3i\varphi_a(-Q_1^a+Q_2^a+Q_3^a)}
\ee
\no and
\be
e^{3i\varphi_b(-Q_1^b + Q_2^b + Q_3^b)}
\ee
\end{subequations}
\no The invariance of the interaction (3.3) therefore implies the
conservation of $Q^a$ and $Q^b$, or of the electric charge and
hypercharge.

By (2.23) and (2.24) $m+m^\prime$ and $m-m^\prime$ are then conserved
as well as $m$ and $m^\prime$ separately, i.e.
\be
\begin{array}{rcl}
m_1 &=& m_2 + m_3 \\
m_1^\prime &=& m_2^\prime + m_3^\prime \\
\end{array}
\ee

\vskip.5cm

\section{Spectrum of the Algebra.}

The operator $b\bar b$ is a self-adjoint operator with real
eigenvalues and orthogonal eigenstates.  It follows from the algebra
that
\be
b\bar b|n\rangle = q^{2n}|\beta|^2|n\rangle
\ee
So that $b\bar b$ resembles the Hamiltonian of an oscillator but
with eigenvalues arranged in geometric progression and with 
$|\beta|^2$ corresponding to $\frac{1}{2}\hbar w$.

$\bar a$ and $a$ are raising and lowering operators respectively:
\begin{eqnarray}
\bar a|n\rangle &=& \lambda_n|n+1\rangle \\
a|n\rangle &=& \mu_n|n-1\rangle
\end{eqnarray}
\no If the $n$ are normalized
\be
\langle n|m\rangle = \delta(n,m)
\ee
\no then
\begin{eqnarray}
|\lambda_n|^2 &=& 1-q^{2n}|\beta|^2 \\
|\mu_n|^2 &=& 1-q^{2(n-1)}|\beta|^2
\end{eqnarray}
\no If $q$ is real as we assume, then there are no finite representations of this algebra obtained by imposing
\begin{eqnarray}
& &\bar a|M\rangle = 0 \\
& &a|M^\prime\rangle = 0 \\
& &M > M^\prime
\end{eqnarray}
\no since these relations are then inconsistent.

One may obtain a finite representation, however, by imposing
\[
\bar a|M\rangle = 0
\]
\no to cut off the spectrum at the top level and by imposing a physical
boundary condition at the bottom level by interpreting $|0\rangle$
as the state of lowest energy.  We may interpret $|n\rangle$ as the
state with $n$ nodes.  This physical boundary condition is obviously
externally imposed and supplements the algebra.

\vskip.5cm

\section{Phenomenlogy.}

In standard electroweak theory the masses of the vector bosons are
determined by the interaction of the vector and Higgs fields, while
the mass of the Higgs itself is fixed by the Higgs potential.  In the
same theory the masses of the fermions are provisionally associated
with a term of the following form
\be
\bar\psi_L\varphi \psi_R + \bar\psi_R \bar\varphi\psi_L
\ee
\no where $\psi$ and $\varphi$ are the fermionic and Higgs fields
respectively and where $\psi_L$ and $\varphi$ are isotopic doublets
while $\psi_R$ is an isotopic singlet.

Without essentially altering the Lagrangian of the standard model
one could replace (5.1) by
\be
\bar\psi_L\varphi V(\bar\varphi\varphi)\psi_R +
\bar\psi_R V(\bar\varphi\varphi)\bar\varphi\psi_L
\ee
\no where $V(\bar\varphi\varphi)$ is the Higgs potential, which
determines the mass of the Higgs, or alternatively where
$V(\bar\varphi\varphi)$ may introduce different interaction terms
that determine the masses of the fermions.

We shall take $\psi_R$ to be a singlet as in the standard theory.  We
shall also assume that all quantum fields, including the ``Higgs" are
defined over the algebra, and we shall additionally assume that the
potential $V(\bar\varphi\varphi)$, determining the masses of the
fermions has minima at the four points occupied by the four
trefoils.  These points are labelled by the four monomials in (2.2)
and will be referred to as trefoil points.  At these points the 
scalar $\varphi$ and the spinor $\psi$ of the Lorentz group have the
same representation in the internal algebra.

Then (5.2) reduces at the trefoil points to
\be
2\bar\varphi\varphi V(\bar\varphi\varphi) = F(\bar\varphi\varphi)
\ee
\no where
\be
\varphi(w,r) \sim D^{3/2}_{\frac{w}{2}\frac{r+1}{2}}
\ee
\no Hence the mass operator is a functional of
\be
\bar\varphi\varphi\sim \bar D^{3/2}_{\frac{w}{2}\frac{r+1}{2}}
D^{3/2}_{\frac{w}{2}\frac{r+1}{2}}
\ee
\no The eigenstates of $\bar\varphi\varphi$ are then the eigenstates
of $b\bar b$ since every $a$ is compensated by an $\bar a$ in the
product (5.5) and therefore $\bar D^{3/2}_{\frac{w}{2}\frac{r+1}{2}}
D^{3/2}_{\frac{w}{2}\frac{r+1}{2}}$ lies in the $(b,\bar b)$ 
subalgebra.  It follows that the eigenvalues $m_n(w,r)$, the masses
of the trefoils, are given by
\be
F\left(\bar D^{3/2}_{\frac{w}{2}\frac{r+1}{2}}
D^{3/2}_{\frac{w}{2}\frac{r+1}{2}}\right)|n\rangle =
m_n(w,r)|n\rangle
\ee
\no depending on $F$.  By (5.4) and (5.6) each of the four trefoils
represented by $(w,r)$ may exist in various excited states $n$.  We
assume that only the lowest three states are occupied.  In the
lepton family, for example, these states (0,1,2) are occupied by
$(e,\mu,\tau)$.  With an allowed choice of $V(\bar\varphi\varphi)$ one may
obtain a finite mass spectrum for the fermions by imposing the
algebraic boundary condition $\bar a|M\rangle=0$ at the top level
and by also imposing a physical boundary condition at the bottom
level by interpreting $|0\rangle$ as the state of lowest energy.

Eq. (5.6) permits one to calculate relative masses of the twelve
fermions.  Ratios of masses in the same family may be computed
without ambiguity.  These calculations depend on either
experimental or theoretical knowledge of $V(\bar\varphi\varphi)$.
If $V=1$, one has the standard model and the earlier results.$^{1,2}$

It is also possible to compute reaction rates by taking (3.3)
between definite particle states as follows:
\begin{eqnarray*}
& &\langle n|\bar D^{3/2}_{m_1m_1^\prime} D^1_{m_2m_2^\prime}
D^{3/2}_{m_3m_3^\prime}|n^\prime\rangle \\
& &=\langle n|\bar\varphi(w_1r_1) W(w_2r_2)\varphi(w_3r_3)|n^\prime\rangle
\end{eqnarray*}
\no where
\be
\begin{array}{rcl}
n &=& 0,1,2 \\
n^\prime &=& 0,1,2
\end{array}
\ee

In this way one may calculate relative masses of fermions by (5.6)
and relative reaction rates mediated by vector bosons between
fermions by (5.7).  This work has been carried out in some detail
in Ref. 2 but the model may be refined as more empirical information
is utilized.

\vskip.5cm

\no {\bf References.}
\begin{enumerate}
\item R. J. Finkelstein, Int. J. Mod. Phys. A{\bf 20}, 487 (2005).
\item R. J. Finkelstein and A. C. Cadavid, hep-th/0507022.
\end{enumerate}

\end{document}